\documentclass[referee]{aa}

\usepackage{graphicx}
\usepackage{natbib}
\usepackage{latexsym}
\usepackage{amsmath}
\usepackage{amssymb}
\usepackage{amstext}
\usepackage{color}
\usepackage{psfrag}

\def\k{km s$^{-1}$}
\def\ks{km s$^{-1}$~}
\def\d{$^\circ$}
\def\m{$^\prime$}
\def\s{$^{\prime\prime}$}
\def\hh{$^{\mathrm h}$}
\def\mm{$^{\mathrm m}$}
\def\ss{$^{\mathrm s}$}
\def\cm3{cm$^{-3}$}

\def\12{$^{12}$CO}

\begin{document}

\title{High resolution CO observations towards the Bright Eastern Knot of the SNR Puppis A}

\author {S. Paron \inst{1}
 \footnote[1]{Post-Doctoral Fellow of CONICET, Argentina} 
 \and G. Dubner  \inst{1}
 \footnotemark[2]
  \fnmsep
 \footnote[3]{Visiting Astronomer, European Southern Observatory}
 \and E. Reynoso \inst{1,2} 
 \footnote[2]{Member of the Carrera del Investigador Cient\'\i fico of CONICET, Argentina}
 \and M. Rubio \inst{3}
}

\institute{Instituto de Astronom\'{\i}a y F\'{\i}sica del Espacio (IAFE),
             CC 67, Suc. 28, 1428 Buenos Aires, Argentina\\
             \email{sparon@iafe.uba.ar} 
\and Departamento de F\'\i sica, Facultad de Ciencias Exactas y Naturales, UBA, Buenos Aires, Argentina
\and Departamento de Astronom\'\i a, Universidad de Chile, Casilla 36-D, Santiago, Chile} 

\offprints{S. Paron}

   \date{Received <date>; Accepted <date>}

\abstract{}{This paper reports molecular observations towards the Bright Eastern Knot (BEK) in the 
SNR Puppis A, a feature where radio and X-ray studies suggest that the shock front is interacting 
with a dense molecular clump.}{We performed high-resolution millimetric observations towards the BEK of Puppis A 
using the SEST telescope in the \12 J=1--0 and 2--1 lines (beams of 45\s~and 23\s~respectively). 
More extended, lower angular resolution \12 J=1--0 observations taken from NANTEN archival data
were also analyzed to obtain a complete picture.}{ 
In the velocity range near 16 \k, the Puppis A systemic velocity, our study revealed two important properties: 
(i) no dense molecular gas is detected immediately adjacent to the eastern border of the BEK and (ii) the 
molecular clump detected very close to the radiocontinuum maximum is probably located in the foreground along the line 
of sight and has not yet been reached by the SNR 
shock front. We propose two possible scenarios to explain the absence of molecular emission eastwards of the BEK border 
of Puppis A. Either the shock front has completely engulfed and destroyed a molecular clump or the 
shock front is interacting with part of a larger cloud and we do not detect CO emission immediately beyond it
because the molecules have been dissociated by photodissociation and 
by reactions with photoionized material due to the radiative precursor.}{}

\titlerunning{High resolution CO observations towards the Puppis A BEK}
\authorrunning{S. Paron et al.}


\maketitle

\section{Introduction}

Shock waves generated by supernova remnants (SNRs) can accelerate, compress,
heat, fragment or even destroy surrounding interstellar clouds. Strong shock-cloud interactions can
enhance or reduce abundances of different molecular species with respect to quiescent cloud conditions. 
Observations of molecular gas associated with SNRs provide
information essential to understand the physics and chemistry involved in these processes.

Puppis A is a Galactic SNR that has been extensively studied in the whole electromagnetic 
spectrum. In radio continuum, Puppis A appears as an asymmetric clumpy shell with 
the brightest section along the eastern border (Figure \ref{observations}a) (\citealt{gabi06} and references therein) 
presenting a good
correlation with soft X-ray emission \citep{petre82,aschen88}. The X-ray emission includes both extended 
features and compact knots,  
the most conspicuous of which is the bright eastern knot (BEK; \citealt{petre82}) that coincides with 
an indentation in the shock front of the SNR as seen in radio. Such morphology suggests an interaction 
between the SNR and a dense interstellar clump.  
\citet{hwang05} presented ACIS {\it Chandra} X-ray images and spectral data of the region 
around the BEK. They conclude that a cloud-shock 
interaction in a relatively late phase of evolution is taking place near the BEK, while closer to the forward shock in 
the BEK region, the SNR has recently interacted with a more dense and extended obstacle. 

CO studies of the interstellar medium surrounding Puppis A performed by 
\citet{dubner88} with angular resolution of 8\m.7, 
revealed the existence of a chain of molecular clouds 
concentrated along the E and NE periphery of the remnant. Studies in infrared wavelengths
of Puppis A also suggest the existence of interstellar clouds along the eastern border of the SNR
\citep{arendt91}.
From the atomic and molecular studies \citep{dubner88,reyno95}, a systemic velocity 
of $v_{LSR} \sim +16$ \ks and a kinematical distance of $2.2 \pm 0.3$ kpc were derived for Puppis A, 
a distance later confirmed by \citet{reyno03} based on interferometric, high resolution HI data 
in the direction of the associated radio quiet neutron star 
RX J0822-4300 located within the remnant.

In this work, we present high
resolution SEST observations performed in the \12 J=1--0 and J=2--1 lines towards the BEK region, complemented 
with lower resolution molecular mapping of a larger region including the BEK based on archival data of the 
NANTEN telescope.
The posibility of interaction between the SNR shock front and the surrounding gas is locally investigated 
in an attempt to understand the origin of the enhancement observed both in X-ray and radio continuum emissions.

\section{Observations}

The high-resolution \12 data were acquired during March 9 to 11, 2000 with the 15 m Swedish-ESO Submillimetre Telescope 
(SEST) that operated in La Silla (Chile). 
The angular resolutions of this telescope were 45\s~and 23\s~for the \12 J=1--0 
and J=2--1 transitions respectively.
An Acousto-Optical spectrometer was used as back end, consisting of a
narrow band high-resolution (HRS) spectrometer with 1000 channels, bandwidth 80 MHz,
and channel separation 41.7 kHz (corresponding to 0.108 \ks for the \12 J=1--0 and 0.054 \ks for the \12 J=2--1). 
The observed velocity ranges were [$-$40 \k,$+$60 \k] and [$-$10 \k,$+$40 \k] for the \12 J=1--0 and J=2--1 
transitions respectively.

\begin{figure*}[tt]
\centering
   \includegraphics[width=12cm]{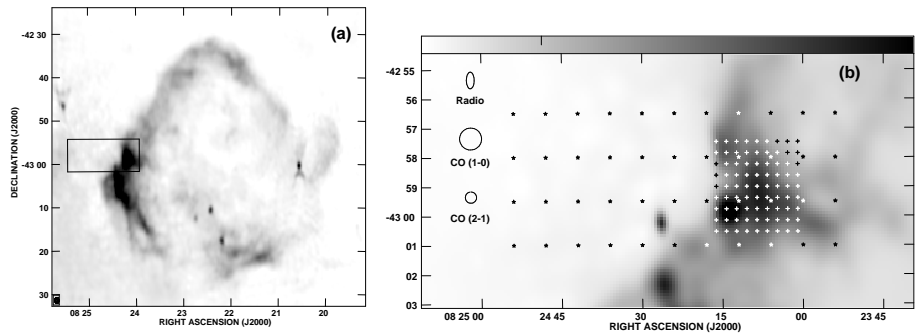}
     \caption{{\bf (a)} Radio continuum image of Puppis A at 1425 MHz (from \citealt{gabi06}). The rectangle 
shows the region surveyed with SEST in the \12 J=1--0 and J=2--1 lines.
{\bf (b)} Observed region in detail. The pointings are indicated with crosses and stars. 
The respective beams are shown to the left. }
     \label{observations}
\end{figure*}

Figure \ref{observations}a displays the SNR Puppis A as observed in radio continuum at 1420 MHz. 
The rectangle shows the region surveyed with the SEST telescope. In this region 
the \12 J=1--0 and 2--1 transitions were observed in 81 pointings with a grid spacing of 23\s.
The pointings are shown as crosses in the enlargement included in Figure \ref{observations}b.  
This region covers a square of approximately 
3\m.35 $\times$ 3\m.35 centered at RA $=$ 8\hh 24\mm 8.5\ss, dec $=-$42\d 59\m 00\s~(J2000). The whole 
field was observed twice with an integration time of 105 sec per position each time.
Beyond this square, 44 additional pointings, also shown in Figure \ref{observations}b as stars, were observed in 
both CO lines with a grid spacing of 90\s, except for one pointing 
(at RA $=$ 8\hh 23\mm 54\ss~and dec $= -$43\d 01\m 00\s) that was observed only at the \12 J=1--0 
transition.

All spectra were Hanning smoothed to improve the signal-to-noise ratio.
The rms noises in main-beam brightness temperature are $\sigma_{1-0} \sim 0.15$ K and $\sigma_{2-1} \sim 0.30$ K for 
the 81 pointings observed over the BEK  and $\sigma_{1-0} \sim 0.20$ K and $\sigma_{2-1} \sim 0.15$ K for the 
44 additional pointings for the \12 J=1--0 and 2--1 transitions respectively.
The spectra were processed using the XSpec software package developed at the Onsala Space Observatory. Images were
produced using the AIPS package.

Additionally we used \12 J=1--0 data acquired with the 4 m radiotelescope NANTEN that operated in Las Campanas Observatory 
(Chile), to explore a more extended region along the eastern border of Puppis A. These data are part of the
the study in the direction of the Gum Nebula carried out by \citet{yama99}. 
The angular resolution of the telescope is 2\m .6 and the grid sampling is 8\m~in l and b. The total 
bandwidth and the effective spectral resolution were 40 MHz and 35 kHz respectively, corresponding to a velocity 
coverage of 100 \ks and a velocity resolution of $\sim$ 0.1 \k. At this velocity resolution the typical rms noise
was $\sim$ 0.7 K \citep{yama99}.

\section{Results}

\begin{figure}[h]
\centering
   \includegraphics[width=4cm,angle=-90]{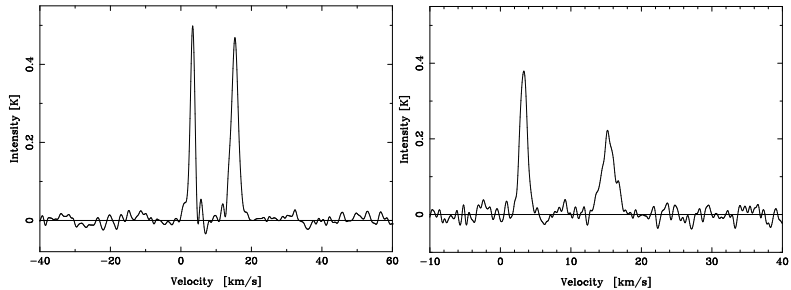}
     \caption{\12 J=1--0 ({\it left}) and J=2--1 ({\it right}) average profiles from the 81 observed spectra 
     towards the BEK of Puppis A (crosses in Figure \ref{observations}b).}
     \label{ave}
\end{figure}

Figure \ref{ave} shows \12 J=1--0 and J=2--1 spectra in the whole observed velocity range obtained from the 
average of the 81 pointings towards the BEK of Puppis A. Two velocity components are present, near 
3 and 16 \k. 
The narrow 3 \ks component probably originates in foreground unperturbed gas crossed by the line of sight,
likely to be associated with the Gum Nebula \citep{reyno97,yama99}. This component will therefore not be
further considered.

\begin{figure}[h]
\centering
\includegraphics[width=10cm]{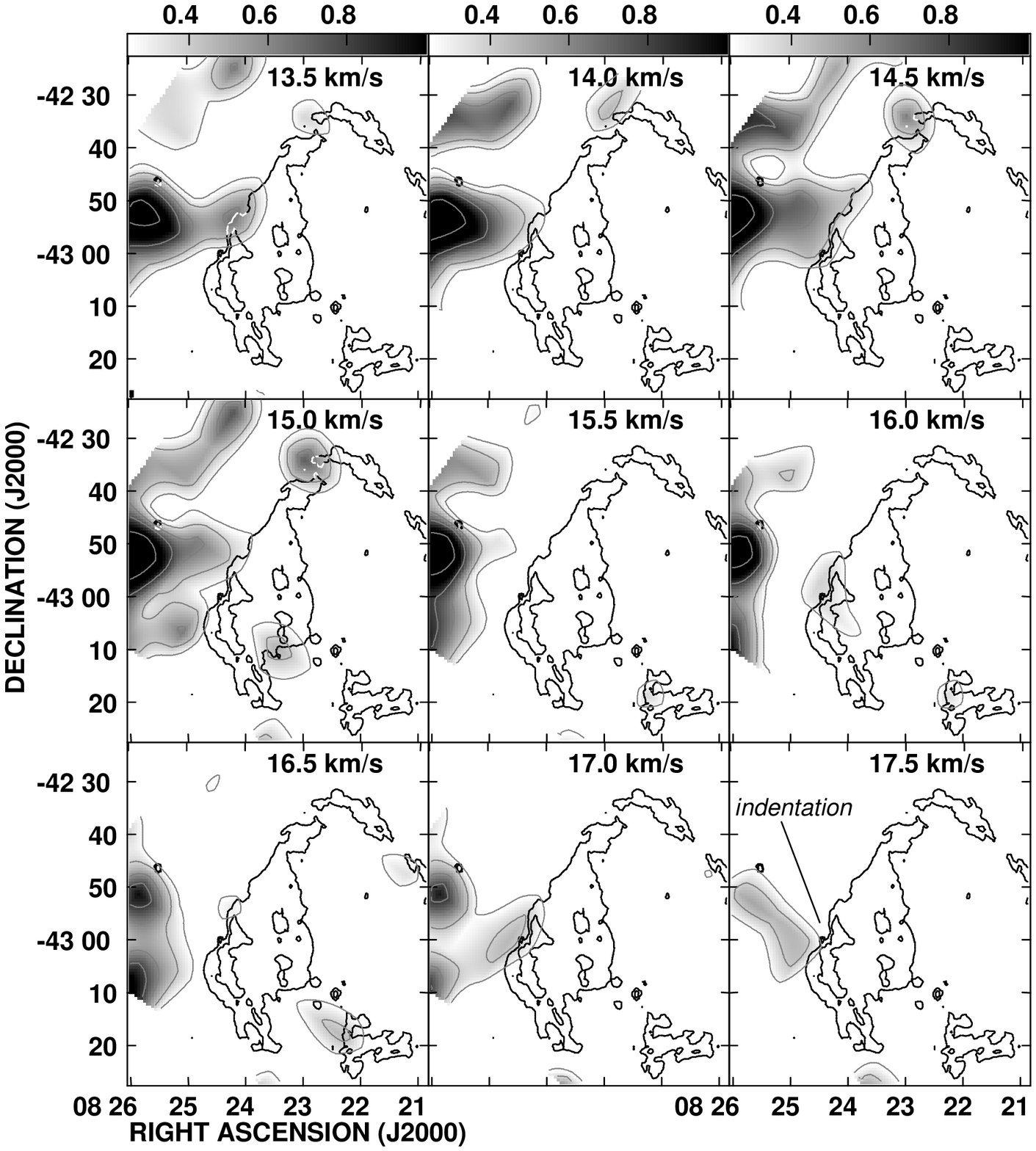}
\caption{\12 J=1--0 emission obtained with NANTEN (from \citealt{yama99}) integrated each 0.5 \k.
Some contours of the radio continuum emission of Puppis A are included. In the last frame,
the position of the BEK indentation is indicated. The gray scale, in K \k, is shown at the top. The contour levels of
the \12 emission are 0.25, 0.40, 0.60, 0.80 and 1.20 K \k. The rms noise of each image is $\sim 0.06$ K \ks and the 
beam is 2\m.6.}
\label{nanten}
\end{figure}

To analyze the surroundings of the BEK area we first inspect the distribution of the molecular gas near $v \sim 16$ \ks 
in an extended region (1\d $\times$ 1\d) 
around Puppis A based on the coarser sampled NANTEN observations. In Figure \ref{nanten} we present the \12 J=1--0 
emission integrated each 0.5 \ks in the range between 13.5 and 17.5 \k. The rms noise, calculated from regions free of 
emission, is 0.06 K \k.
The NANTEN data confirm the presence of molecular gas eastwards of the SNR 
around $v \sim 16$ \ks as suggested before based on lower angular resolution observations \citep{dubner88}. 
The maximum of the molecular complex, however, is shifted to the east in
at least half a degree from the border of the SNR. Therefore, it is likely that the SN shock expanding towards 
the east encountered moderately dense molecular gas, maybe with some embedded denser clumps that cannot be 
resolved with these observations.
The morphological matching between the edges of the SNR and the molecular complex between 
13.5 and 15.0 \ks is striking, but
the \12 J=1--0 emission observed between $v = 15.5$ and 16.5 \ks  is clearly detached from the SNR's eastern 
limb. At $v = 16.0$ \k, in coincidence with the systemic velocity derived for Puppis A, a very faint
molecular feature 
can be seen exactly covering the region of the BEK. In summary from the analysis of these moderate resolution data,
in spite of the proximity of the eastern cloud to the SNR limb, it is not obvious
that the SN shock is abutting upon a dense molecular cloud. In what follows, we investigate the BEK region with 
greater detail using the SEST data. 

\begin{figure}[h]
\centering
\includegraphics[width=4.82cm,angle=-90]{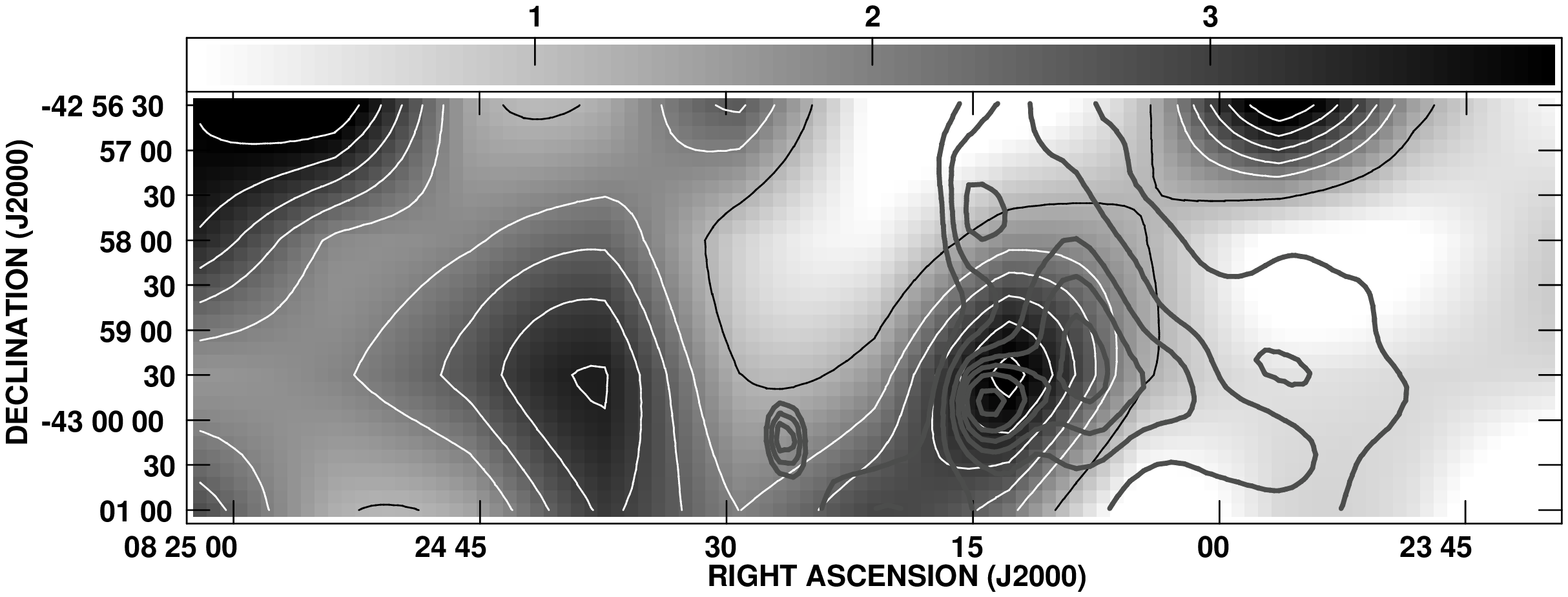}
\vspace{0.4cm}
\includegraphics[width=4.4cm,angle=-90]{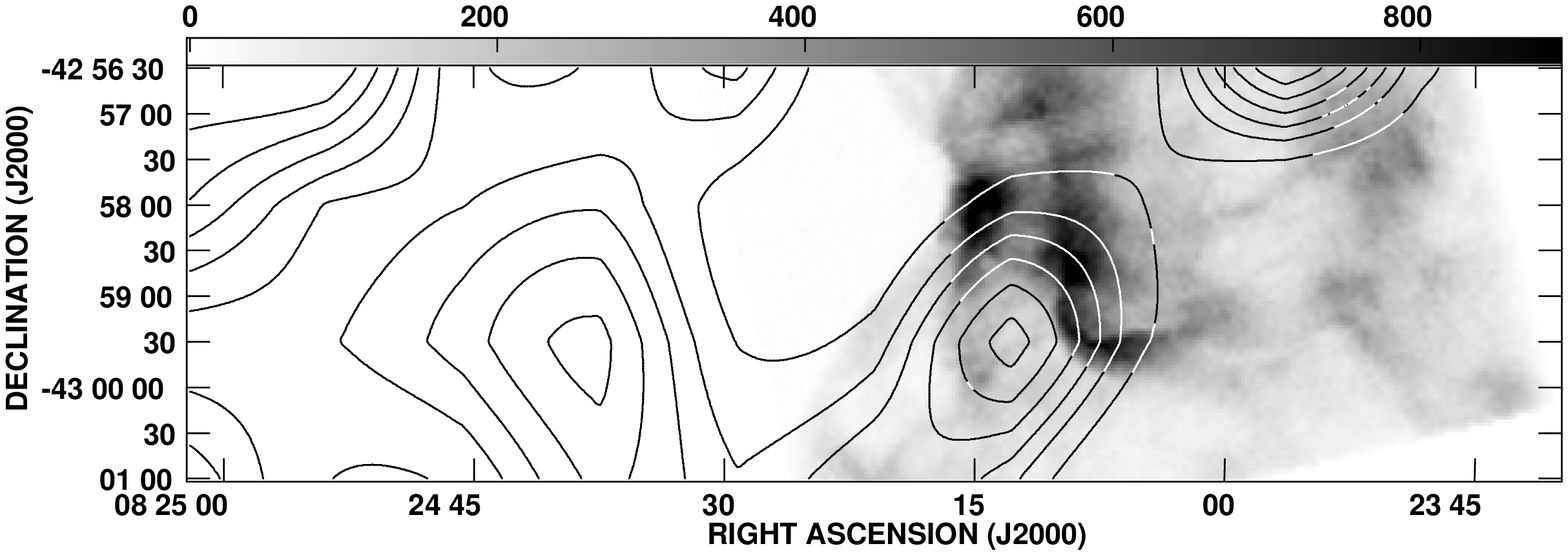}
\caption{  {\it Top panel}: \12 J=1--0 emission, integrated between 13.5 and 17.5 \ks
obtained from the 44 pointings observed with a grid spacing of 90\s, is presented in grays and contours
with levels 1.2, 2, 2.5, 3, 3.5, 4 K \k. The thick, dark grey contours correspond to the radio continuum emission
of Puppis A with levels 23, 36, 50, 60, 68, 76, 87 mJy beam$^{-1}$. {\it Bottom panel}: the same \12 J=1--0
contours presented in the top panel, superimposed over the {\it Chandra} X-ray emission
(image taken from the {\it Chandra} SNR Catalog).}

\label{integ16sparRadio}
\end{figure}

Figure \ref{integ16sparRadio} displays
the \12 J=1--0 integrated between 13.5 and 17.5 \ks obtained from the SEST 44 pointings performed with
a grid spacing of 90\s~(see Figure \ref{observations}) compared with the radio continuum emission (top) and with 
the {\it Chandra} X-ray image (bottom). 
A molecular clump very close to the radio continuum maximum and south of the brightest 
X-ray features, is detected. Clearly detached from this CO emission, there is a second concentration towards 
the east, 
probably part of the complex shown in Figure \ref{nanten}, and another feature, not completely mapped, 
towards the northwest of the mapped region, well inside the SNR 
shell. These observations confirm that no high density CO gas is observed
immediately adjacent to the eastern border in this part of Puppis A in spite of the radiocontinuum and X-ray
emission morphologies, suggestive of higher compression in this region due to the shock front encountering
a dense molecular cloud.

\begin{figure}[h]
\centering
   \includegraphics[width=13cm]{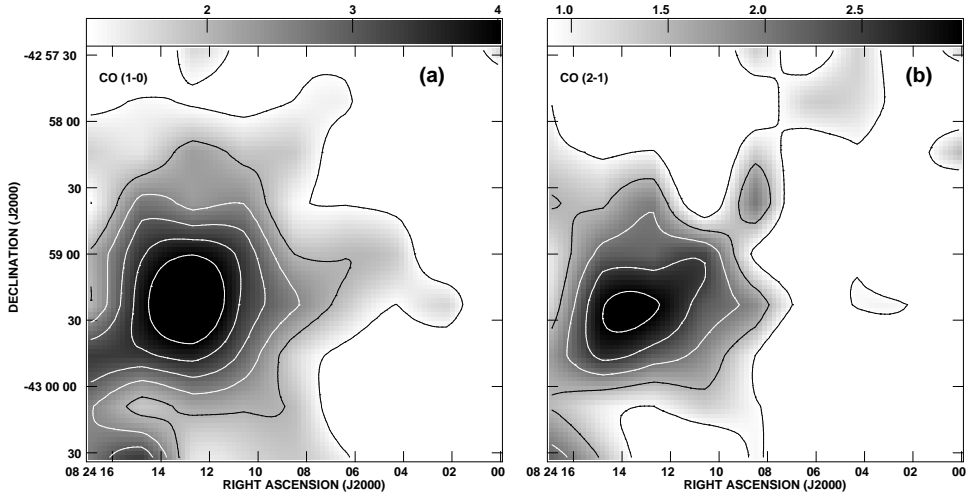}
\caption{\12 emission obtained from the 81 pointings observed towards the BEK. {\bf (a)} \12 J=1--0 emission
integrated between 13.5 and 17.5 \k. The contour levels are 1.2, 2, 2.5, 3, 3.5
and 4 K \k. The beam size is 45\s~and the noise level
is $\sigma \sim 0.4$ K \k. {\bf (b)} \12 J=2--1 emission integrated between 13.5 and 17.5 \k. The contour levels
are 0.9, 1.5, 2, 2.5 and 3 K \k. The beam size is 23\s~and the noise level
is $\sigma \sim 0.3$ K \k. In both images the gray scale is shown at the top.}
     \label{integ16}
\end{figure}

\begin{figure}[h]
\centering
   \includegraphics[width=13cm]{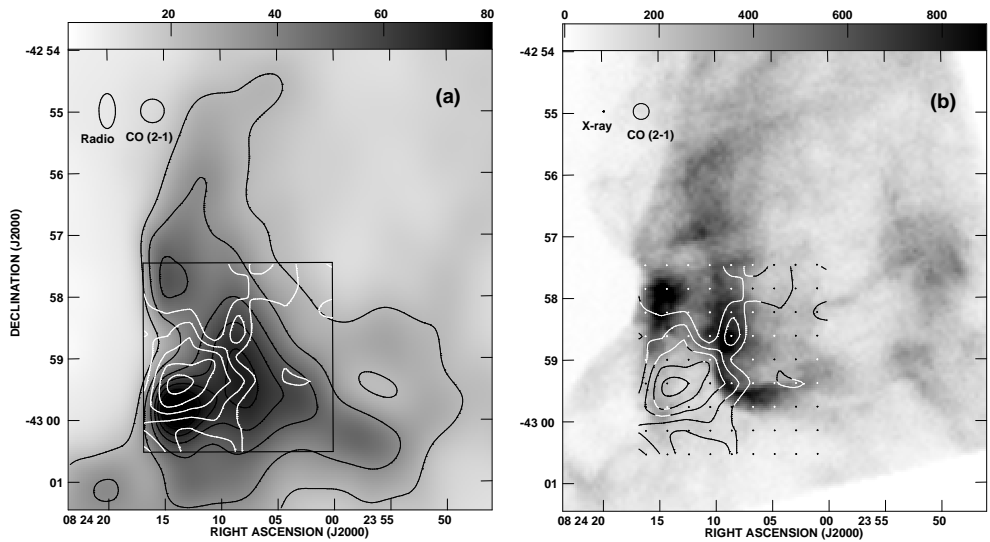}
 \caption{   {\bf (a)} \12 J=2--1 contours over the radio continuum
emission of Puppis A, the levels are the same of Figure 5b. Also some contours (in black)
of the radio continuum are included with levels 23, 36, 50, 60, 68, 76, 87 mJy beam$^{-1}$.
The gray scale, in mJy beam$^{-1}$, is shown at the top.
{\bf (b)} Same \12 J=2--1 contours over the {\it Chandra} X-ray emission (from the {\it Chandra} SNR
Catalog). The points indicate the pointings for the molecular observations. The respective beams are shown at the top left
corner of each image.}
\label{co+cont}
\end{figure}

To analize the molecular clump observed very close to the radicontinuum maximum, in
Figure \ref{integ16} we display the \12 J=1--0 and J=2--1 distribution integrated between 13.5 and 17.5 \k,
obtained from the 81 pointings performed with the best grid spacing.
This figure shows with more detail the molecular clump described above. Taking into account that the \12 J=2--1
line is optically thinner and surveys denser gas than the J=1--0 line \citep[e.g.][]{saka94}, we will use it to
compare with the radio continuum and X-ray emissions. In Figure \ref{co+cont}, the same \12 J=2--1 contours shown 
in Figure \ref{integ16}b are displayed over the radio continuum emission (Figure \ref{co+cont}a),
which appears in greys and with black contours, and over the {\it Chandra} X-ray emission (Figure \ref{co+cont}b).
These images show that the maximum of the molecular clump is located spatially very close to, but not exactly coincident
with the maximum of the radio continuum.
At the same time the clump is clearly anticorrelated with the X-ray emission.
There is, however some molecular/X-ray coincidence: the western border of
the clump appears superimposed onto the arched bright 
X-ray feature called ``the bar'' by \citet{hwang05}.

In what follows we calculate the physical parameters of the only molecular clump spatially and kinematically
coincident with the SNR bright eastern radio knot.
If we consider the limit of the \12 J=1--0 structure observed in Figure \ref{integ16}a as 
the 3.5 K \ks contour ($\sim 8$ times above the noise level), a well delimited region of $\sim$1\m~diameter is
defined. Assuming a distance of 2.2 kpc, the linear diameter of this region is 0.3 pc.
Using the \12 J=1--0 integrated emission and the conversion factor X = 1.9 $\times 10^{20}$ cm$^{-2}$ (K \k)$^{-1}$ 
\citep{bloemen86} we can estimate the molecular mass of this feature to be M $\simeq 9.5$ M$_{\odot}$ and its 
volumetric density, $n \simeq 1.7 \times 10^{4}$ cm$^{-3}$.

\section{Discussion}                                  


\citet{hwang05} analyzed in detail the X-ray emission morphology near the region of the BEK of Puppis A,
reporting the detection of two different morphological components in the area. In particular these authors
point out that the indentation immediately east of the X-ray emission peak, strongly suggests
that the shock front has recently interacted with a dense obstacle and is wrapping it around. They conclude 
that this obstacle is probably quite extended along the line of sight and projection effects must be important.
The presence of multiple clouds in the line of sight is a possibility previously noted by \citet{blair95}. 
What is surprising is that the new high-resolution millimetric observations revealed that there is no molecular 
gas in a region immediately adjacent to the eastern border of the BEK.
To explain the lack of dense molecular gas in a fringe about $\sim$3\m~wide (about $\sim$2 pc at 2.2 kpc) to the 
east of Puppis A, in spite of the fact that the presence of a molecular complex close to the 
eastern border is well proved, two possible reasons can be proposed. One of them is that 
the shock front has completely engulfed a molecular clump and no remains are left. In this case, the compact 
X-ray emission could be due to evaporation
of the molecular gas as proposed by \citet{petre82} and \citet{teske87}, and the indentation could be marking the 
reestablishment of the shock front following the cloud passage. 
A similar behavior has been observed in another SNR, the Cygnus Loop 
\citep{fesen92}, where H$\alpha$ observations show an isolated emission cloud along the Cygnus Loop's eastern limb 
and an indented shock front beyond the cloud location. \citet{klein94}, assuming a radiative shock, developed an
hydrodynamical model for the interaction of shock waves with interstellar clouds and applied it to the shocked cloud 
in the eastern region of the Cygnus Loop. They found that, as a result of the interaction, the blast wave 
diffracts around the cloud and 
reestablishes ahead of it, in agreement with the observations of \citet{fesen92}. 
Besides, X-ray {\it ROSAT} observations \citep{levenson97,levenson02} show an indentation at this region of the 
Cygnus Loop followed by a compact bright knot interior to the shell, similarly to what 
is observed in Puppis A.

\begin{figure}[h]
\centering
   \includegraphics[width=14cm]{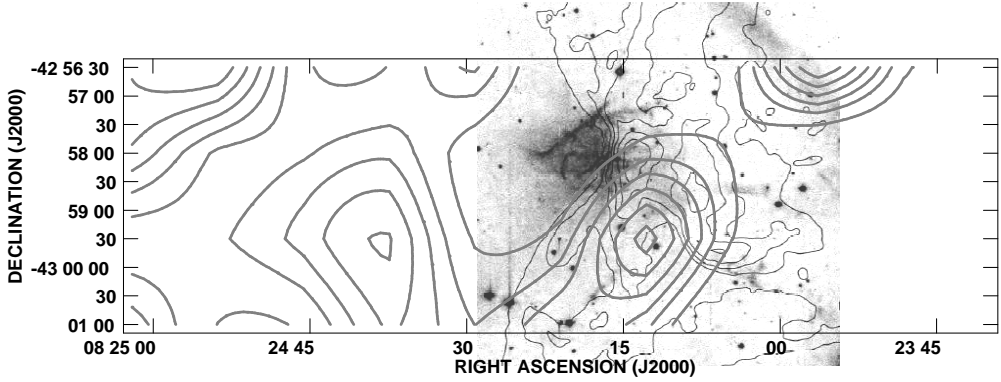}
     \caption{[O III] emitting filaments (observed by \citealt{blair95}) with contours of the X-ray emission
(image extracted from \citealt{hwang05}). The same contours of the \12 J=1--0 emission presented in Figure
\ref{integ16sparRadio} are superimposed.}
\label{oiii}
\end{figure}

An alternative to explain the lack of molecular gas adjacent to the BEK could be that
the indentation is a consequence of 
an interaction with part of a larger cloud and 
we do not detect CO emission ahead of the SNR shock because of photoionization and photodissociation effects. 
Indeed, \citet{blair95} detected [O III] emitting filaments in Puppis A immediately adjacent to the BEK indentation,  
extending significantly beyond the eastern boundary of the X-ray emission. The authors
propose that the enhanced diffuse [O III] to the east could indicate preshock photoionization while the diffuse 
optical emission behind the X-ray shock front may imply evaporation. 
Figure \ref{oiii} shows the [O III] emission image of the region as taken from \citet{blair95}
with contours of the X-ray emission (thin gray lines) and of molecular emission (thick lines) overlaid.
Using the Hopkins Ultraviolet Telescope, \citet{blair95} reported the detection
of far-ultraviolet (FUV) emission in one of the [OIII] filaments near the BEK position 
(at about R.A. $=$ 8\hh 24\mm 18.65\ss~and dec. $= -$42\d 57\m 40\s.50, J2000), which could confirm the existence of 
FUV photons capable of dissociate molecules.
It is known that energetic electrons produced by X-ray ionization in molecular clouds collisionally excite the Lyman
and Werner bands of H$_{2}$; the subsequent radiative de-excitations generate a flux of FUV 
photons capable of dissociating many molecular species including the CO \citep{prasad83,wardle99}. 
It is possible that the CO immediately adjacent to Puppis A's shock front has been 
photodissociated by the FUV photons product of the X-ray ionization. Moreover, reactions
between the CO and He$^{+}$, ionized by the radiative precursor \citep{ghava00}, can dissociate the CO
molecule \citep{prasad83}. 
Far-infrared observations of the dust continuum and the fine structure of [CII] and [OI] at 158 $\mu$m and 63 $\mu$m 
respectively would be helpful to explore the interface between the warm photodissociated region in the cloud's 
surface and the cooler interior \citep{hollen99}. Such detection would provide support to this last scenario.




On the other hand, between $v \sim 13.5$ and $17.5$ \ks we observed a conspicuous molecular structure lying
very close to the maximum
of the radio continuum emission of the BEK. This velocity range coincides with the systemic velocity of Puppis A,
$v \sim 16$ \k, derived by \citet{reyno95,reyno03} and \citet{dubner88}. In order to investigate if this 
clump has been recently shocked,
we calculated the ratio between the integrated lines ($R_{2-1/1-0}$), where
$R_{2-1/1-0} = {\int^{17.5}_{13.5}T_{{\rm CO2-1}}~dv}~/{\int^{17.5}_{13.5}T_{{\rm CO1-0}}~dv}$. This parameter is
a useful tool to trace shocked gas. Large values in the ratio $R_{2-1/1-0}$ are expected to be found in shocked
molecular clouds, as observed
in the clouds related with the SNRs G349.7+0.2 ($R_{2-1/1-0} \sim 1.5$; \citealt{dubner04}), W44 ($R_{2-1/1-0} \sim 1.3$),
IC443 ($R_{2-1/1-0} \sim 1.3-4$) and HB 21 ($R_{2-1/1-0} \sim 1.6-2.3$) \citep{seta98,koo01}.
For the observed molecular clump we obtained $R_{2-1/1-0} \sim 0.6$, a low value that suggests that it is quiescent 
gas.
Figure \ref{co+cont}b shows that the molecular clump is surrounded by regions of enhanced X-ray emission. 
This can be an evidence that the molecular cloud is located in the foreground and absorbs the X-ray emission
from the SNR. A similar behavior is observed in a ``molecular arm'' towards the SNR CTB 109, where the CO emission is 
anticorrelated with the X-ray emission \citep{sasaki06} and interpreted in the same way. 
As mentioned above, based on X-ray spectral data, \citet{hwang05} showed that the BEK region closer to the forward 
shock has lower temperatures and ionization ages and higher column densities than 
elsewhere in the field, concluding that these properties  
are signatures of a recent interaction between the SNR shock front with multiple clouds, 
some of which are in the remnant foreground.
In this context, the discovered molecular structure must be one 
of these foreground clumps that has not been completely reached by the SN shock yet.

\section{Summary and conclusions}

The present observations enabled us to map in detail the molecular material towards the 
brightest X-ray emission region (BEK) in Puppis A.
Our \12 J=1--0 and J=2--1 high resolution observations show components at $v \sim 3$ \ks and $v \sim 16$ \k.
The $v \sim 3$ \ks component is most probably related to the Gum Nebula. 

In the velocity range near 16 \k, our study revealed two important properties: (i) no
dense molecular gas is detected immediately adjacent to the eastern border of the BEK of 
Puppis A (towards the SNR indentation) and (ii) the only detected cloud apparently 
associated with the BEK is probably located in the foreground along the line of sight and 
has not been reached by the SN shock yet. 

To explain the absence of molecular emission eastwards of Puppis A in the BEK region, two scenarios
were proposed. Either the shock front has completely
engulfed and destroyed a molecular clump and the X-ray emission is due to the evaporation of the molecular gas, 
or the indentation is the consequence of an interaction with part of a larger cloud and we do not detect CO 
emission immediately beyond it because the molecules have been dissociated by photodissociation and by reactions
with photoionized material due to the radiative precursor,
as optical emission suggests. Far-infrared observations of the dust continuum and the fine structure of [CII] and 
[OI] may help to test this suggestion.

Concerning the molecular clump detected at $v \sim 16$ \k,
we estimated for it a linear size of about 0.3 pc,  
a molecular mass of $\sim 9.5$ M$_{\odot}$, and a volume density of $\sim 1.7 
\times 10^{4}$ cm$^{-3}$. The low value obtained in the ratio between the integrated lines towards this clump, 
$R_{2-1/1-0} \sim 0.6$, suggests that we are observing quiescent gas located in front of Puppis A along the line 
of sight.

\begin{acknowledgements}
We thank Norikazu Mizuno for kindly providing us with the NANTEN data.
G.D. is very grateful to the staff of SEST for the support received 
during the observations, especially to the former director Dr. L.-A. Nyman.
This work was supported by the CONICET grant 6433/05, UBACYT A055/04 and ANPCYT PICT 04-14018. 
\end{acknowledgements}

\bibliographystyle{aa}  
\bibliography{bib-puppis}
\IfFileExists{\jobname.bbl}{}
{\typeout{}
\typeout{****************************************************}
\typeout{****************************************************}
\typeout{** Please run "bibtex \jobname" to optain}
\typeout{** the bibliography and then re-run LaTeX}
\typeout{** twice to fix the references!}
\typeout{****************************************************}
\typeout{****************************************************}
\typeout{}
}

\end{document}